\documentclass[reprint,amsmath,amssymb,
 aps,
 prl,
 superscriptaddress,twocolumn
]{revtex4-1}

\usepackage{graphicx}
\usepackage{dcolumn}
\usepackage{bm}
\usepackage{color}

\begin{document}

\title{Direct Measurement of Strain-dependent Solid Surface Stress}

\author{Qin Xu}
\thanks{These authors contributed equally to this work.}
\affiliation{Department of Materials, ETH Z\"urich, 8093 Z\"urich, Switzerland.}
\affiliation{Department of Mechanical Engineering and Materials Science, \\ Yale University, New Haven, CT 06511, USA.}

\author{Katharine E. Jensen}
\thanks{These authors contributed equally to this work.}
\affiliation{Department of Materials, ETH Z\"urich, 8093 Z\"urich, Switzerland.}
\affiliation{Department of Mechanical Engineering and Materials Science, \\ Yale University, New Haven, CT 06511, USA.}

\author{Rostislav Boltyanskiy}
\affiliation{Department of Mechanical Engineering and Materials Science, \\ Yale University, New Haven, CT 06511, USA.}

\author{Rapha\"el Sarfati}
\affiliation{Department of Mechanical Engineering and Materials Science, \\ Yale University, New Haven, CT 06511, USA.}

\author{Robert W. Style} \email{robert.style@mat.ethz.ch}
\affiliation{Department of Materials, ETH Z\"urich, 8093 Z\"urich, Switzerland.}
\affiliation{Mathematical Institute, University of Oxford, Oxford, OX1 3LB, United Kingdom.}

\author{Eric R. Dufresne}
\email{eric.dufresne@mat.ethz.ch}
\affiliation{Department of Materials, ETH Z\"urich, 8093 Z\"urich, Switzerland.}

\date{\today}

\begin{abstract}
Surface stress, also known as surface tension, is a fundamental material property of any interface. 
However, measurements of solid surface stress in traditional engineering materials, such as metals and oxides, have proven to be very challenging.
Consequently, our understanding relies heavily on untested theories, especially regarding the strain dependence of this property.
Here, we take advantage of the high compliance and large elastic deformability of a soft polymer gel to directly measure solid surface stress as a function of  strain.
As anticipated by theoretical work for metals, we find that the surface stress depends on the strain via a surface modulus. 
Remarkably, the surface modulus of our soft gels is many times larger than the zero-strain surface tension.
This suggests that surfaces stresses can play a dominant role in solid mechanics at much larger length scales than previously anticipated.
\end{abstract}

\maketitle

All material surfaces are characterised by a surface energy and a surface stress. 
The difference between these material properties is important: the surface energy, $\gamma$, is a scalar equal to the minimum work per unit area to cut a solid, while the surface stress, $\Upsilon_{ij}$, is a tensor that describes the in-plane force per unit length required to stretch a surface.
In simple liquids, the surface stress and surface energy have the same magnitude ($\Upsilon_{ij} = \gamma\delta_{ij}$) and are independent of any deformation.
In solids, both quantities are expected to be strain-dependent.  
They are related through the Shuttleworth equation \cite{shut50},
\begin{equation}
\Upsilon_{ij}=\gamma \delta_{ij}+\frac{\partial \gamma}{\partial \epsilon^s_{ij}},
\label{eq:shutt}
\end{equation}
\noindent where $\delta_{ij}$ is the identity tensor and $\epsilon^s_{ij}$ is the surface strain.
For nearly sixty years, Equation \ref{eq:shutt} has served as the foundation for a well-established body of theory and computation including rigorous analyses of how surface stress can be incorporated into physical models  \cite{gurt75,spae00}, predictions for $\Upsilon_{ij}(\epsilon_{ij}^s)$ in a variety of materials \cite{vand87,gumb91,shen05}, and an extensive literature anticipating the role of strain-dependent surface stresses across various phenomena in nanostructures and nanocomposites  \cite{mill00,ding05,duan05,shar04,he08,lu14}.
Apart from some indications that $\Upsilon_{ij}\neq \gamma \delta_{ij}$ (\textit{e.g.} \cite{dieh12,wolf12}), there is sparse direct experimental evidence to support these theories.
In particular, we are unaware of any experiments that have been able to  measure strain-dependent surface stresses in a solid material \cite{camm94}.

\begin{figure}[t]
\centering 
\includegraphics[width= 0.49\textwidth]{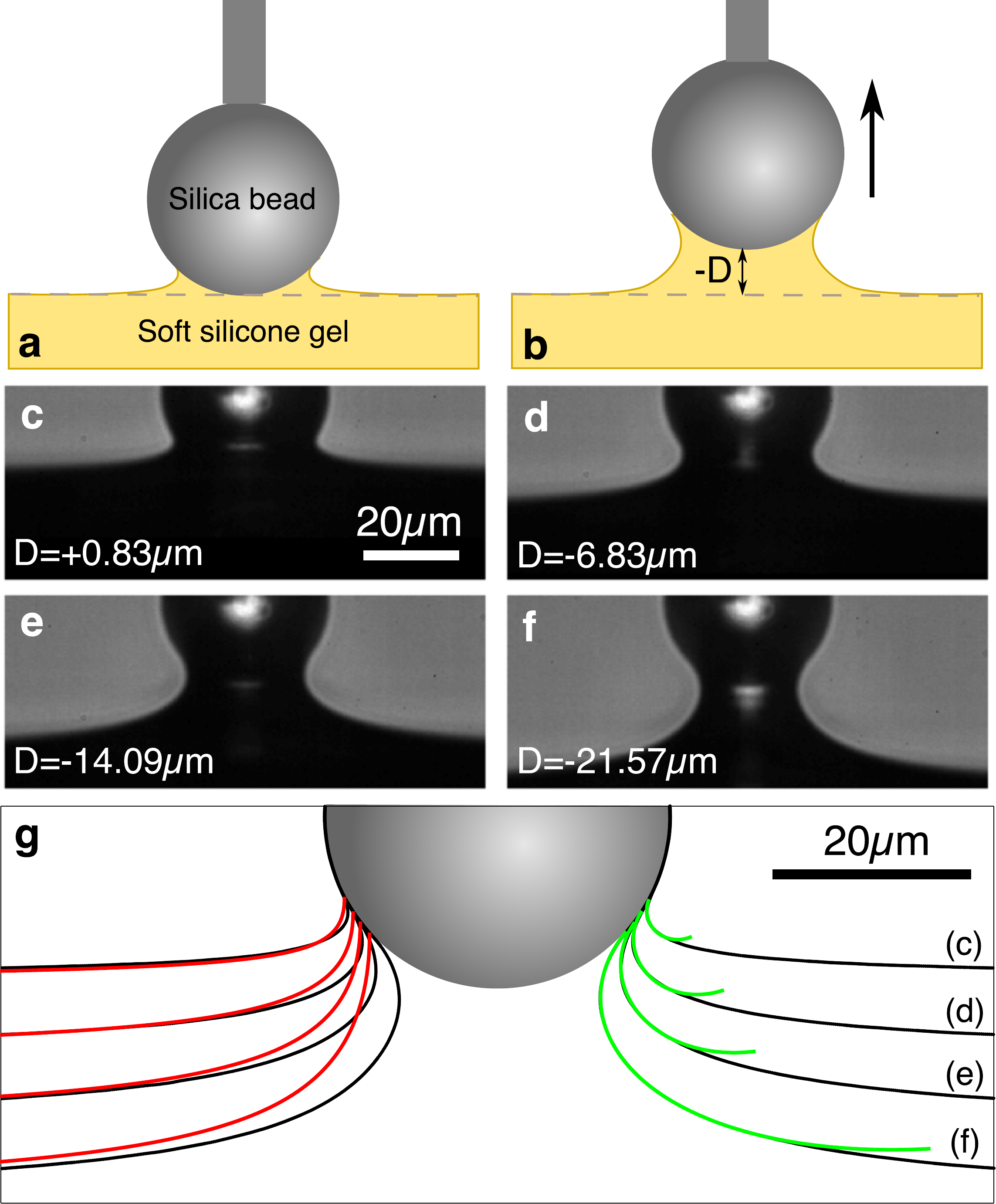}
\caption{{\bf A solid meniscus in soft adhesion.}
(\textbf{a}-\textbf{b}) Schematic of the experiment at (a) initial contact and (b) during quasi-static pull.
(\textbf{c}-\textbf{f}) Raw snapshots of a 17.4-$\mu$m-radius sphere adhered to an initially-flat, compliant ($E = 5.6$ kPa) silicone gel substrate as it is pulled quasi-statically from (c) first contact to (f) the last measured stable position.  
(\textbf{g}) Mapped deformation profiles (black points) corresponding to (c-f),
with predictions of classic elastic theory \cite{john71,maug95} overlaid at left (red lines),
and best-fit constant total curvature surfaces overlaid at right (green lines). 
}
\label{Fig1}
\end{figure}

Here, we directly measure surface stress as a function of strain in a compliant polymer gel.
In such materials, surface stresses can be important at the micron-scale and can be directly measured  from the structure of a three-phase contact line using a light microscope \cite{styl13}.
 In the range of accessible strains (up to 25\%), surface stresses increase linearly.
The effect is not subtle: surface stress doubles with only 17\% strain.

To clearly visualize the impact of strain-dependent surface stress, 
we first image extreme deformations of a soft solid. 
We bring rigid glass spheres with radii from 7.9 to 32.1 $\mu$m into adhesive contact with $\sim$300-$\mu$m-thick, compliant, silicone gel substrates (Young's modulus $E=5.6$ kPa), and quasi-statically retract them, as shown schematically in Fig. \ref{Fig1}a-b.
Below a critical displacement, a  solid, axisymmetric bridge of silicone stably connects the spheres to the substrates, as shown in the example optical micrographs of Fig. \ref{Fig1}c-f.

As we pull on such an adhesive contact, the solid silicone bridge between substrate and sphere takes on a remarkable shape, resembling a liquid meniscus.
This not only differs from the predictions of classic linear elastic theory \cite{john71,maug95}, shown as red lines at left in Fig. \ref{Fig1}g, but also from recent  
large-deformation simulations of Neo-Hookean solids \cite{liu16b}. 
Alternatively, we test the correspondence with a liquid meniscus by fitting the profiles to surfaces of constant total curvature, shown as green curves at right in Fig. \ref{Fig1}g.
The curves capture the shape of the solid free surface near the contact line.
In this example,
at first contact (Fig. \ref{Fig1}c) the total curvature is -0.29 $\mu$m$^{-1}$.
As the bead is retracted, the magnitude of the curvature drops, decreasing to -0.01 $\mu \mathrm{m}^{-1}$ at the last recorded stable position (Fig. \ref{Fig1}f).

Why should the free surface assume a liquid-like shape near the contact line?
Recent theory and experiment, reviewed in \cite{styl16}, have found that surface stresses dominate bulk elastic stresses at wavelengths smaller than a characteristic elastocapillary length scale, $\Upsilon/E$ \cite{styl16}.
Thus, we  expect a capillary-dominated near field within this distance of the three-phase contact line.
The size of the domain of constant curvature at initial contact in Fig. \ref{Fig1}c is 5.1 $\mu$m, defined as the path length over which the capillary solution fits the measured profile.
Indeed, this is comparable to the previously-measured elastocapillary length $\Upsilon/E$ for this material, about 4$~\mu \mathrm{m}$   \cite{jens15}.
However, the size of the constant curvature domain does not remain constant with deformation.
Rather, it increases dramatically with sphere displacement, reaching 31.8 $\mu$m at the last stable position 
in Fig. \ref{Fig1}f.

The growth of this solid meniscus 
with increasing strain suggests a proportional increase in the elastocapillary length, $\Upsilon/E$.
This can only be due to an increase in the surface stress with tensile strain, as Young's modulus is constant up to about 10\% strain and then increases slightly thereafter (see Supplementary Information).
This suggests a more than six-fold increase in $\Upsilon$ during the deformation in Fig. \ref{Fig1}. 
However, this experimental geometry is not well-suited to a quantitative measurement of the relationship between surface stress and strain.
The strain in the solid meniscus is highly inhomogeneous, and the size of the domain of constant curvature depends on both the bulk and surface mechanical properties.  

\begin{figure*}[t]
\centering 
\includegraphics[width=\textwidth]{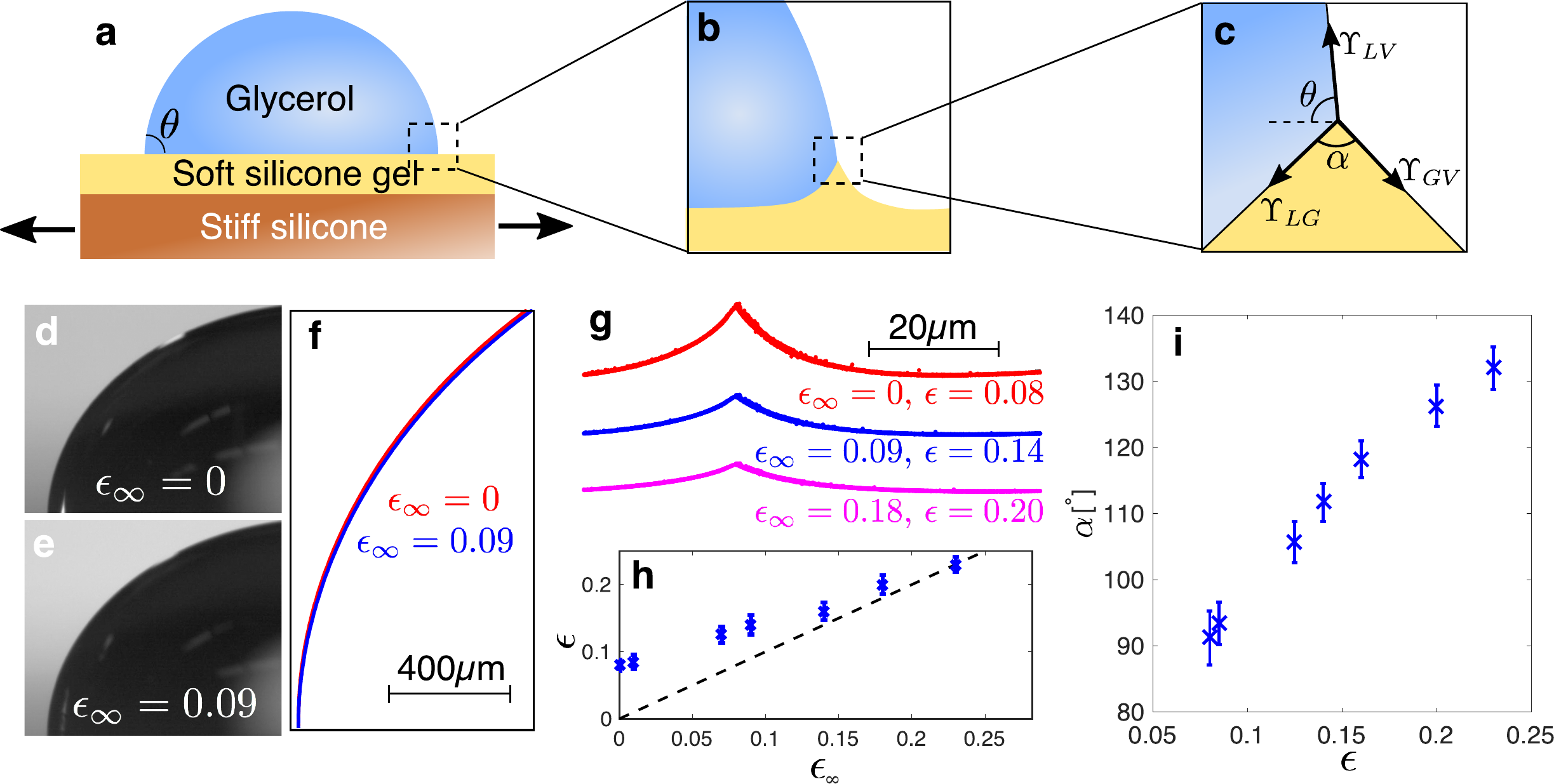}
\caption{{\bf Macroscopic contact angle and microscopic wetting profiles.} 
({\bf a}) Schematic of the strain-dependent wetting experiments, using a biaxial stretcher as described in Ref. \cite{na08}. 
({\bf b}) Detail of the contact line geometry at intermediate scales
({\bf c}) Detail of the contact line at microscopic scales, much less than $\Upsilon/E$.  At this scale, the geometry of the contact line is given by a vector balance of the surface stresses as shown. 
({\bf d}) \& ({\bf e}) Macroscopic wetting profiles of large glycerol droplets sitting on unstretched and stretched ($\epsilon_{\infty}=0.09$) silicone gels. 
({\bf f}) Superimposed boundaries for the drops on the stretched (blue) and unstretched (red) substrates show no difference in the macroscopic contact line geometry.   
({\bf g}) Microscopic wetting profiles for a single droplet on unstretched (red), $9\%$ stretched (blue), and $18\%$ stretched (pink) silicone gel substrates, respectively. 
({\bf h}) Local strain near the contact point, $\epsilon$, plotted against the applied strain, $\epsilon_\infty$. Dashed line has a slope of 1. 
({\bf i}) The opening angle  of the wetting ridge, $\alpha$, increases with the local strain, $\epsilon$. In (H) and (I), the error bars are standard deviations of the population.}
\label{Fig2}
\end{figure*}

Motivated by these observations, we designed
wetting experiments that allow us to measure directly the local relationship between surface stress and strain. 
We measure the macroscopic and microscopic structure of the contact line of glycerol droplets on soft silicone substrates ($E=3.0~\mathrm{kPa}$) as we apply a uniform biaxial strain, $\epsilon_\infty$, as shown schematically in Fig. \ref{Fig2}a-c (see Supplementary Information for details).
While the macroscopic contact angle for large droplets depends only on the surface energies (Fig. \ref{Fig2}a) \cite{styl12},  the microscopic structure at the contact line with such a soft substrate is governed by a balance of surface stresses (Fig. \ref{Fig2}c) \cite{styl13,park14,bost14,cao15b}.

Macroscopic measurements on large droplets show no change of the contact angle with applied strain; it always re-equilibrates to its original value of $\theta=90.8^\circ$ following a stretch (Fig. \ref{Fig2}d-f).
This demonstrates that there is no significant contact-line hysteresis (see Supplementary Information), and that the surface energies of the solid-liquid and solid-vapor interfaces are nearly identical.  

The microscopic structure of the contact line, on the other hand, reveals the surface stresses \cite{styl13}.
Within distances from the contact line smaller than $\Upsilon/E$, each of the three intersecting interfaces becomes straight and has an orientation given by the mechanical balance of the surface stresses, as shown in Fig. \ref{Fig2}c.
Importantly, this shape is defined locally and is independent of the bulk elastic properties of the material.
We measure the profile of the silicone substrate near the contact line using confocal microscopy, initially with zero applied strain.
The substrate forms a symmetric wetting ridge at the contact line, as seen in Fig. \ref{Fig2}g.
With no applied strain, the ridge is about 10 $\mu$m high and has an opening angle $\alpha=91.2^\circ$, determined by fitting the region close to the contact line as intersecting lines.

Knowing the shape of the substrate near the contact line,  we can measure the local surface stress by applying the force balance of Fig. \ref{Fig2}c.
Since the macroscopic contact angle is nearly $90^\circ$, we know that the liquid-air interface divides the opening angle $\alpha$ nearly equally, and the horizontal force balance reduces to $\Upsilon_{LG}\approx\Upsilon_{GV}=\Upsilon$ \cite{jeri11}. 
Balancing the surface stresses along the vertical axis further requires that $\Upsilon=\Upsilon_{LV}/(2\cos(\alpha/2)) $
where $\Upsilon_{LV}$ is the liquid-vapor surface tension, which we measured to be $41\pm 1~\mathrm{mN/m}$ for glycerol droplets in contact with the silicone substrate (see Supplementary Information). 
The result is a measured solid surface stress of $\Upsilon=29$ mN/m, about 50\% larger than the surface tension of silicone liquids and consistent with our previous measurements using a similar experimental geometry \cite{styl13}. 

In marked contrast to the macroscopic measurements, the microscopic contact line geometry changes significantly when we stretch the substrate.
The contact line geometry for the same droplet at  different values of applied biaxial strain, $\epsilon_\infty$, is shown in Fig. \ref{Fig2}g.
To ensure that the contact line has reached equilibrium, we wait at least 40 minutes after applying each strain. 
As the applied strain increases from 0 to 18\%, the ridge height decreases by a factor of three and the opening angle increases to $\alpha = 126.3^\circ$.  
This change in the opening angle indicates that the surface stress has increased to 44 mN/m, an approximately 50\% increase from the value at $\epsilon_\infty = 0 $.

It is important to note that the strain at the contact line is a combination of the macroscopically-applied strain, $\epsilon_\infty$, and the localised deformation that produces the wetting ridge.
To meaningfully interpret the change in surface stress with applied strain, we measured the  strain at the contact line, $\epsilon$, as a function of the applied strain, $\epsilon_\infty$ (as described in the Supplementary Information).  
Even when the applied strain is zero ($\epsilon_{\infty}=0$), the substrate is stretched in the wetting ridge with a local value of $\epsilon=8\%$ at the contact line. 
With increasing stretch, the wetting ridge flattens out and $\epsilon$ converges toward $\epsilon_{\infty}$, as shown in Fig. \ref{Fig2}h.

\begin{figure}[t]  
\centering 
\includegraphics[width=0.49\textwidth]{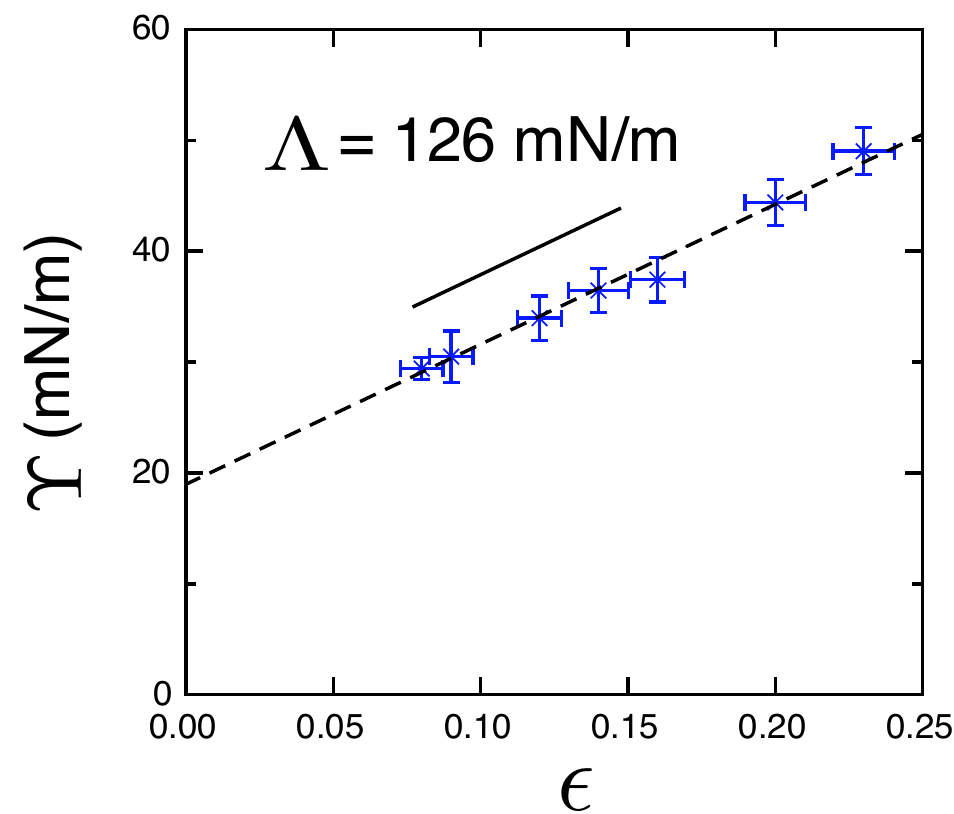}
\caption{{\bf Strain dependence of solid surface stress, $\Upsilon$}. The points indicate the average surface stress and average local strain for droplets on the same substrate.  The error bars are the standard deviation of the population. The dashed line is a linear fit, providing the surface modulus, $\Lambda=126 \pm 6$ mN/m, and zero-strain surface stress, $\Upsilon_0=19 \pm 3$ mN/m.}
\label{Fig3}
\end{figure}

Armed with the ability to extract local measures of both the surface stress and surface strain, 
we measured the wetting ridge profiles of 35 droplets having radii from 12 to 220 $\mu \mathrm{m}$ at seven different externally-applied strains from 0 to 23\%.
As expected, the capillary-dominated structure near the contact line was identical for all droplets under the same strain conditions,
even though the far-field profiles were elasticity-dominated and depended on the droplet size  (Supplementary Fig. 5).
We plot the measured opening angle, $\alpha$, versus strain, $\epsilon$, for all experiments in Fig. \ref{Fig2}i.
As shown in Supplementary Fig. 5, the size of the  capillary-dominated domain increases with applied strain, mirroring our observations from the adhesion experiment in Fig. \ref{Fig1}g.

Applying the force balance of Fig. \ref{Fig2}c to the data in Fig. \ref{Fig2}i, we calculate the strain dependence of the solid surface stress, shown in Fig. \ref{Fig3}.
Over the range of measured strains, up to about 25\%, we find that the surface stress increases linearly with strain. 
Fitting these data to the form $\Upsilon=\Upsilon_0+\Lambda\epsilon$, we find $\Upsilon_0=19 \pm 3$ mN/m and $\Lambda = 126\pm 6$ mN/m.
The fitted value of the surface stress at zero strain, $\Upsilon_0$, is quite close to the surface tension of the gel's liquid silicone precursor, which we measured to be $21$ mN/m (see Supplementary Information). 

The surface modulus, $\Lambda$, is a material property of the solid silicone gel surface.
Surface moduli are predicted to play an important role in the nanoscale mechanics of metallic surfaces, but have never been measured directly \cite{ibac97,shar04,he08}.
As with bulk elasticity, elastic surfaces can be characterised by two surface moduli that represent the response of the surface to shear or biaxial stretch: the surface equivalents of shear and bulk moduli \cite{styl16}.
In our experiments, the strain is nearly isotropic, so $\Lambda$ is nearly equal to the latter.

The strain dependence of the surface stress is remarkably strong, with $\Lambda \gg \Upsilon_0$.
Consequently, the surface stress increases $2.5\times$ with 25\% strain. 
This dramatic effect emphasizes just how important strain-dependent solid surface stress is to a complete description of the mechanics of compliant materials. 
It may also resolve mystery in the reported values of the surface stress of soft silicones, which have thus far varied from 19 to 70 mN/m, since each measurement has involved a different specific experimental geometry and thus different strain states (see summary in Supplementary Information)  \cite{jens15,styl13c,park14,jago12,nade13,mond15}.

Decades of theoretical work have been based on the expectation that strain-dependent solid surface stress is a universal feature of all solids, but direct measurements in conventional stiff materials have proven extremely challenging. 
Soft solids, with their high compliance and ability to sustain large-strain elastic deformation, provide unique experimental model systems that make accessible direct measurements of this fundamental material property.
Moving forward, we anticipate a rich interplay between experiments and theory for a broad range of solid materials. 
For example, experimental studies probing elastocapillary mechanics in micro-structured polymer gels and elastomers \cite{mora10,jago12,nade13,ducl14,styl15} could inform our understanding and design of nano-structured metals and semiconductors \cite{chen06,he08,lu14}.

Numerous applications already rely on compliant solids, from adhesives to soft robotics to medical implants \cite{cret03b,kim13,mine15,moon03,rose14}.
Our results have immediate consequences for elastocapillary phenomena in these materials \cite{andr16,andr16b,chak13,styl13c,styl13b,styl15,gonz15,karp16b}.
Thanks to the strong strain dependence of the surface stress,  the elastocapillary length scale can be tuned 
and significantly extended
through careful control of the stress state.
This suggests an exciting new design space where adhesion and wetting properties of a soft material could be modulated with mechanical stimuli. 
To fully realize these ideas, a fundamental investigation of the structure-property relationships that determine the surface modulus in polymeric materials is required.  
We expect that much can be learned from comparison to complex fluid-fluid interfaces, where a sophisticated understanding is emerging of the structure-property relationships that underlie surface rheology \cite{full12,herm15}.

\vskip 0.2in
\noindent \textbf{Acknowledgments:} We thank Dominic Vella, Jay Humphrey, Anand Jagota, Herbert Hui, Frans Spaepen, and Bob Cammarata for useful discussions.
We thank the NSF (CBET-1236086) and ARO MURI W911NF-14-1-0403 for support. \\

\bibliography{stybib2}

\end{document}